\documentstyle[12pt,psfig,preprint,aps]{revtex}

\begin{document}
\title{Azimuthal correlations of pions in relativistic heavy ion collisions at
1 GeV/nucl.}
\author{
        S. A. Bass$^{a,b}$, C. Hartnack$^{b,c}$, H.~St\"ocker$^a$ and
W.~Greiner$^a$}
\address{
        $^a$Institut f\"ur Theoretische Physik der J. W. Goethe
Universit\"at\\
	$^{~}$Postfach 11 19 32, 60054 Frankfurt am Main, Germany\\
        $^b$GSI Darmstadt, Postfach 11 05 52, 64220 Darmstadt, Germany\\
        $^c$Laboratoire de Physique  Nucl\'{e}aire, Nantes, France
}

\maketitle

\begin{abstract}
Triple differential cross sections of pions
in heavy ion collisions at 1 GeV/nucl.
are studied with the IQMD model.
After discussing general properties of $\Delta$ resonance and pion
production  we focus on azimuthal correlations:
At projectile- and target-rapidities we observe an anticorrelation in the
in-plane transverse momentum between pions and protons.
At c.m.-rapidity, however, we find that
high $p_t$ pions are being preferentially emitted perpendicular to
the event-plane.
We investigate the causes of those correlations and their sensitivity
on the density and momentum dependence of
the real and imaginary part of the nucleon and pion optical potential.
\end{abstract}

\pacs{25.75+r}

\pagebreak

\section{Introduction}
One of the main goals of the study of relativistic heavy ion collisions is the
determination of the
density and momentum dependence of the real and imaginary part of the nucleon
(and other hadron)
optical potential (often also termed nuclear equation of state)
\cite{sch68,cse86,sto86,st86,cl86,schue87,cas90}.
Its importance stretches well beyond nuclear physics
and is of great importance for the formation of nuclear matter
after the big bang, the behaviour of supernovae and neutron stars.
It also is important for the quest for  the quark-gluon plasma in heavy ion
collisions.

An increasing number of observables which are accessible through
heavy ion collisions has been found to be sensitive to the equation of
state: Among the most prominent ones are collective flow effects such as
the bounce--off of cold spectator matter {\em in} the reaction plane
\cite{st80} and the
squeeze--out of hot and compressed participant matter {\em perpendicular}
 to the reaction plane \cite{st82} as well as particle production
\cite{st78,da79,st81}.
The pion multiplicity was one of the
first observables suggested to be sensitive to the nuclear equation of
state \cite{st78,da79,st81}.
This was motivating a strong experimental effort ($4 \pi$ analysis of
streamer chamber events at the BEVALAC) \cite{san80,sto82,har85}.
However, the sensitivity of pion yields and spectra \cite{nag81}
on the equation of state
is not very high \cite{bert84a,kru85a}
and therefore the attention shifted towards {\em subthreshold}
production of mesons (e.g. kaons and $\eta$-mesons)
\cite{shor,ai85b,ai87b,berg94,mis94}.

New experimental $4\pi$ setups at two of the major heavy ion reasearch
facilities, GSI (FOPI, KaoS, TAPS) and LBL (TPC), enable the
investigation of the emission pattern and correlations of
primary and secondary particles in a
far more detailed manner than ever before.
It is now for the first time possible
to thoroughly investigate correlation phenomena such as in--plane bounce--off
\cite{ha88,lib91a,baplb,tra94} and
out--of--plane squeeze--out \cite{baprl,brill93a,ven93}
of {\em pions}. The detailed investigation of these effects, including
their possible origin and their impact parameter and $p_t$ dependencies
as well as their sensitivity to the nuclear equation of state, are the
subject of this publication.

\section{The IQMD-Model}
The first widely used microscopic models for the description of relativistic
heavy ion collisions were based on the Vlasov--Uehling--Uhlenbeck (VUU)
theory \cite{kru85a,moli85b,ai85a},
which explicitly treats nonequilibrium and (stochastic) quantum effects in
the framework of one--particle quantities,
as well as the nuclear potential (nuclear equation of state).
The dynamical basis of the VUU--model is the following transport equation:
\begin{eqnarray*}
   \frac{\partial f}{\partial t} &+&
   \vec{v} \cdot \nabla_r f - \nabla_r U \cdot \nabla_p f \quad
 = \quad - \frac{4}{(2\pi)^3} \int {\rm d}^3 p_2 \, {\rm d}^3 p_1 ' \,
                                {\rm d}\Omega \,
     v_{12} \frac{{\rm d}\sigma}{{\rm d}\Omega} \\ \nonumber
   &\times& \left [ f f_2 (1-f_1 ') (1-f_2 ') -f_1 ' f_2 ' (1-f)
   (1-f_2) \right ] \\
&\times&\delta^3 (p+p_2 -p_1 ' - p_2 ' ).
\end{eqnarray*}
$f$ is the single--particle distribution function.
The l.h.s. contains the potential $U$. Usually $U$ is parameterized
using  the Skyrme ansatz. This gives the opportunity to study the effects of
the nuclear equation of state via different parameter sets.

The r.h.s. contains the cross section $\sigma$ and the
Nordheim--Uehling--Uhlenbeck modifications
incorporating the Pauli--blocking factors
\cite{ue33}.
Models based on the same theory, but differing in numerical implementation,
are the Boltzmann--Uehling-Uhlenbeck (BUU) \cite{ai85a} and the
Landau--Vlasov \cite{greg87} model.
All 3D-numerical implementations of the VUU theory are solved
with the test particle
method. The number of test particles used to represent a nucleon
varies with the
numerical implementation. The test particle method
solvs Hamilton's equation of motion for each test particle.
These transport models have been successful in studying various aspects
of relativistic heavy ion collisions, such as single particle spectra,
collective effects (stopping, bounce--off, squeeze--out) and
meson production.

However, certain fluctuations and correlations,
such as the formation of fragments in
relativistic heavy ion collisions, cannot be studied with a transport model
based on a single--particle distribution function.
This was one of the motivations
for the developement of
the {\bf Q}uantum {\bf M}olecular
{\bf D}ynamics model (QMD) \cite{ai87b,ai86,pei89,ai91}.
In the QMD model the baryons are represented by Gaussian
shaped density distributions
\begin{displaymath}
 f_i (\vec{r},\vec{p},t) = \frac{1}{\pi^2 \hbar^2 }
 e^{-(\vec{r} - \vec{r}_{i0} (t) )^2  \frac{1}{2L} }
 e^{-(\vec{p} - \vec{p}_{i0} (t) )^2  \frac{2L}{\hbar^2}  }
\end{displaymath}
They are initialized in a sphere of a radius
$R=1.14A^{1/3}$ fm, in accord with the liquid drop model. Each nucleon
occupies a volume of $h^3$, so that phasespace is uniformly
filled. The initial momenta are randomly choosen between 0 and the local
Thomas-Fermi-momentum. The $A_P$ and $A_T$ nucleons interact via two-
and three- body skyrme forces, a Yukawa potential and
momentum dependent interactions.
Subsequently, the FMD \cite{fmd}, AMD\cite{amd} and PQMD\cite{pqmd}
models have been developed to offer
an improved treatment of the Pauli-principle.

Isospin is treated explicitely (in the socalled ``I''QMD version),
a symmetry potential
(to achieve corrected distributions of protons and neutrons in the nucleus)
and explicit Coulomb forces between the $Z_P$ and $Z_T$ protons are included.

Pion production is treated via the delta resonance
\cite{ha88,ha89,hart}. A frozen $\Delta$ approximation
(infinite lifetime for the $\Delta$ resonance) had
been used  in other versions of
the QMD model.

The hadrons are propagating under the influence of the potential in
Hamilton's equations of motion:
\begin{displaymath}
\dot{p}_i = - \frac{\partial H} {\partial q_i} \qquad
\dot{q}_i =   \frac{\partial H} {\partial p_i}
\end{displaymath}
with
\begin{displaymath}
H=T+V= \sum_i \frac{p_i^2}{2m_i} +
\sum_{i} \sum_{j>i}
 \int f_i(\vec{r},\vec{p},t) \,
V^{ij}
 f_j(\vec{r}\,',\vec{p}\,',t)\,
d\vec{r}\, d\vec{r}\,'
d\vec{p}\, d\vec{p}\,' \quad.
\end{displaymath}
The baryon-potential $ V^{ij} = V^{ij}_{loc}+V^{ij}_{Yuk}+V^{ij}_{Coul}+
V^{ij}_{mdi}+V^{ij}_{sym} $
consists of
\begin{eqnarray*}
V^{ij}_{loc}&=&
t_{1} \delta (\vec{r}_i - \vec{r}_j) +
\sum_{k>j>i} \int f_k(\vec{r}\,'',\vec{p}\,'')
t_{2} \delta (\vec{r}_i - \vec{r}_j)
\delta (\vec{r}_i - \vec{r}_k) d\vec{r}\,'' d\vec{p}\,'' \\
 V^{ij}_{Yuk}&=&
 t_3\frac{\hbox{exp}\{|\vec{r_i}-\vec{r_j}|/\mu\}}
 {|\vec{r_i}-\vec{r_j}|/\mu}\\
 V^{ij}_{Coul}&=&\frac{Z_i Z_j e^2}
 {|\vec{r_i}-\vec{r_j}|}\\
 V^{ij}_{mdi}&=&
t_4\hbox{ln}^2 (1+t_5(\vec{p}_i-\vec{p}_j)^2)\delta (\vec{r}_i -\vec{r}_j)\\
V^{ij}_{sym}&=& t_6 \frac{1}{\varrho_0}
 T_{3i} T_{3j} \delta(\vec{r}_i - \vec{r}_j)
\end{eqnarray*}
The three-body term in $V^{ij}_{loc}$ as stated above is only valid
for a hard equation of state (for a soft equation of state a VUU-type
formulation $\sim \rho^{\gamma}$ has to be used),
$Z_i,Z_j$ denote the charges of the baryons $i$ and $j$ and $T_{3i},T_{3j}$
are their respective $T_3$ components.
The meson-potential only consists of the Coulomb potential.

The parameters $\mu$ and $t_1 ... t_6$ are adjusted to the real
part of the nucleon optical potential.
For the density dependence of the nucleon optical potential
standard Skyrme type parametrizations are used. Two
different equations of state have been implemented: A
hard equation of state (H) with a compressibility of 380 MeV and a
soft equation of state (S) with a compressibility of
200 MeV \cite{kru85a,moli85b}.
A fit of the momentum dependence to measurements \cite{ar82,pa67}
of the real part of the nucleon optical potential \cite{schue87,ai87b,bert88b}
yields:
\begin{displaymath}
        \delta \cdot \mbox{ln}^2 \left( \varepsilon \cdot
                \left( \Delta \vec{p} \right)^2 +1 \right) \cdot
                        \left(\frac{\rho}{\rho_0}\right)
\end{displaymath}
The equation of state in its standard Skyrme type parametrization including
momentum dependence then reads:
\begin{displaymath}
U \,=\, \alpha \cdot \left(\frac{\rho}{\rho_0}\right) +
        \beta \cdot \left(\frac{\rho}{\rho_0}\right)^{\gamma} +
        \delta \cdot \mbox{ln}^2 \left( \varepsilon \cdot
                \left( \Delta \vec{p} \right)^2 +1 \right) \cdot
                        \left(\frac{\rho}{\rho_0}\right)
\end{displaymath}

The mean field notation with the parameters $\alpha, \beta, \gamma, \delta$
and $\epsilon$ has been chosen for reasons of simplicity and in order to
compare the parameters with those used in VUU/BUU calculations. Their values
can be found in table \ref{eostab}. While the forces are calculated
via the nucleon density in VUU/BUU calculations,
a sum over two-particle interactions is performed
in QMD/IQMD calculations.

The parameters $t_1 ... t_6$ are calculated in the IQMD model before the
initialization of the projectile and target nuclei from the
tabulated values of $\alpha, \beta, \gamma, \delta$ and $\epsilon$
which serve as input. The width and normalization of the
Gaussian wave-packets have to be taken into account for the proper
determination of the force-parameters.

Hard N-N-collisions are included by employing
the collision term of the well known VUU/BUU
equation \cite{st86,kru85a,ai85a,wo90,lib91b}.
The collisions are done
stochastically, in a similar way as in the cascade models \cite{yar79,cug80}.
Two particles collide if their minimum distance $d$ fulfills
\begin{displaymath}
 d \le d_0 = \sqrt{ \frac { \sigma_{\rm tot} } {\pi}  }  , \qquad
 \sigma_{\rm tot} = \sigma(\sqrt{s},\hbox{ type} ).
\end{displaymath}
``type'' denotes the ingoing collision partners ($N-N, N-\Delta, N-\pi ...$).
In addition, the Pauli blocking (of the final state) of baryons
is taken into account
by checking the phase space densities in the final states of a two body
collision.
The final phase space fractions $P_1$ and $P_2$ which are already occupied
by other nucleons are determined
for each of the two scattering baryons.
The particular attempt for a collision is then blocked with the probability
\begin{displaymath}
P_{block} \,=\, 1 - (1 - P_1) (1 - P_2)
\end{displaymath}
Whenever an attempted collision is blocked the scattering partners maintain
the
original momenta prior to scattering.
Delta decays are checked in an analogous fashion with
respect to the phase space of the resulting nucleon.

\begin{figure}
\centerline{\psfig{figure=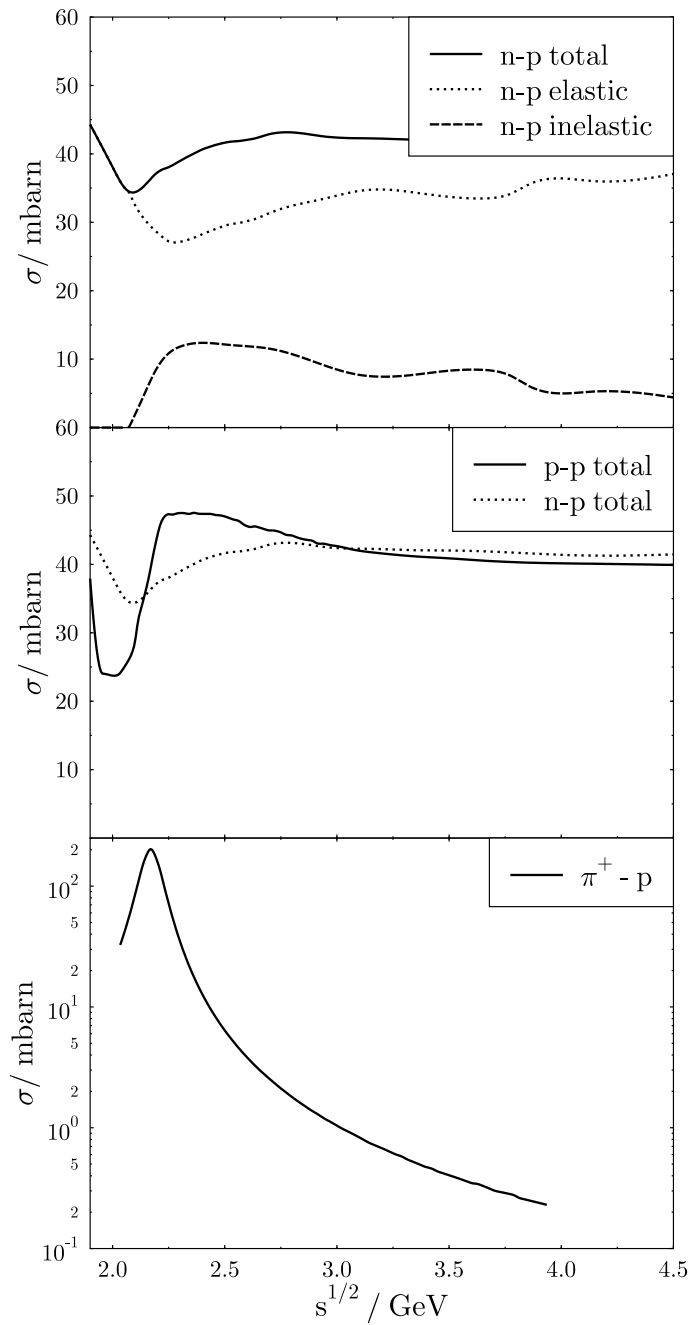,height=18cm}}
\caption{ \label{cross} Tabulated cross sections in the IQMD model. The upper
frame
shows the total, elastic and inelastic proton neutron cross section. The
middle
frame compares the total proton neutron with the total proton proton cross
section
and the lower frame shows the total $\pi^+$ proton (or $\pi^-$ neutron)
cross section. The other
pion nucleon cross sections are determined by scaling the $\pi^+$ proton
cross section
either with $1/3$ ($\pi^+$ neutron and $\pi^-$ proton) or with $2/3$
($\pi^0$ neutron and $\pi^0$ proton). }
\end{figure}

Pions are formed in the IQMD model via the decay of the delta resonance.
The following inelastic reactions are explicitly taken into account
and constitute the imaginary part of the pion optical potential,
which is dominant in the 1 GeV/u  energy domain \cite{eng94}:
\begin{displaymath}
\renewcommand{\arraystretch}{1.2}
\begin{array}{lrcll}
\mbox{a)} & N \, N & \rightarrow & \Delta \, N &
\mbox{ ({\sl hard--delta}--production)}\\
\mbox{b)} & \Delta & \rightarrow & N \,  \pi    & \mbox{ ($\Delta$--decay) }
\\
\mbox{c)} & \Delta \, N & \rightarrow & N \, N &
	\mbox{ ($\Delta$--absorption)}\\
\mbox{d)} & N \, \pi & \rightarrow & \Delta    &
\mbox{ ({\sl soft--delta}--production)}
\end{array}
\end{displaymath}
Elastic $\pi - \pi, \pi - N, \pi - \Delta,
\Delta - \Delta, \Delta - N$ scattering
is not taken into account.
Experimental cross sections are used for processes a) and d) \cite{vw82},
as well as for the elastic N-N-collisions.
The respective cross sections are shown in figure \ref{cross}.

For the delta absorption, process c),
we use a modified detailed balance formula
\cite{da91}. The conventional detailed balance formula is only correct for
particles with infinite life-times (zero width). If the principle of detailed
balance is applied to the delta resonance, then its finite width has to be
taken into account:
\begin{displaymath}
\frac{d \sigma^{N \Delta \rightarrow N N}}{d \Omega} \,=\,
\frac{1}{4} \, \frac{m_{\Delta} p_{NN}^2}{p_{N \Delta}} \,
\frac{d \sigma^{NN \rightarrow N \Delta}}{d \Omega} \,
\left( \frac{1}{2\pi} \int \limits_{m_N + m_{\pi}}^{\sqrt{s} - m_N}
                p_{N \Delta} A_r(M) 2 M \, dM \right)^{-1}
\end{displaymath}
with
\begin{displaymath}
A_r(M) \,=\, \frac{\Gamma / 2 }{( M - m_{\Delta} )^2 + ( \Gamma / 2)^2}
\end{displaymath}
The mass-dependent $\Delta$--decay width has been taken from \cite{rand}:
\begin{displaymath}
\Gamma_{\Delta}(p_{\pi})\,=\, \frac{r(p_{\pi})}{r(p_0)} \, \Gamma_0 \qquad
\mbox{with} \quad r(p)= \frac{p^3}{1 + \left(\frac{p}{p_1}\right)^2
				     + \left(\frac{p}{p_1}\right)^4 } \quad .
\end{displaymath}
$p_{\pi}$ is the decay momentum of the pion, $p_0=227$ MeV/c, $p_1=238$ MeV/c,
$p_2=318$ MeV/c and $\Gamma_0=120$ MeV.
The $\Delta$ decays isotropically in its restframe.

The elastic nucleon--nucleon scattering angular distribution
is taken to be \cite{cug81}:
\begin{displaymath}
\frac{\mbox{d} \sigma_{\mbox{el}}}{\mbox{d}\Omega} \sim \exp(A(s)\;t)\quad,
\end{displaymath}
where $t$ is $-q^2$, the squared momentum transfer and
\begin{displaymath}
A(s)\,=\, 6\, \frac{(3.65\;
(\sqrt{s} - 1.8766))^6}{1 + (3.65 \;(\sqrt{s} - 1.8766))^6}
\quad .
\end{displaymath}
$\sqrt{s}$ is the c.m. energy in GeV and A is given in (GeV/c)$^{-2}$.

The inealstic channel is treated in an analogous fashion.
The  parametrization suggested by Huber and
Aichelin \cite{hub94} is used: fitted differential
cross sections are extracted from OBE calculations:
\begin{displaymath}
\frac{\mbox{d} \sigma_{\mbox{in}}}{\mbox{d}\Omega} \sim a(s) \;
			\exp(b(s)\;\cos \theta)\quad,
\end{displaymath}
$a(s)$ and $b(s)$ are functions of $\sqrt{s}$ and vary in their
definition for different intervals of $\sqrt{s}$ (see table \ref{inelast}).

Pions propagate between collisions (imaginary part of the pion optical
potential) on curved trajectories
with Coulomb forces acting upon them.
The different isospin channels are
taken into account using the respective Clebsch--Gordan--coefficients:
\begin{displaymath}
\renewcommand{\arraystretch}{1.2}
\begin{array}{lcllcl}
\Delta^{++} & \rightarrow & 1( p + \pi^+ ) & \Delta^+    & \rightarrow &
\frac{2}{3} ( p + \pi^0 ) + \frac{1}{3} ( n + \pi^+ ) \\
\Delta^0    & \rightarrow & \frac{2}{3} ( n + \pi^0 ) + \frac{1}{3}
( p + \pi^-) \qquad & \Delta^-    & \rightarrow & 1 ( n + \pi^- ) \\
\end{array}
\end{displaymath}
The real part of the pion optical potential is treated in the following
manner:
As far as the pion is bound in a $\Delta$-resonance, the
density and momentum dependent real part of the nucleon optical
potential is applied as an approximation to the (yet unknown) real part
of the $\Delta$ optical potential. Due to the large $\pi-N$ cross section,
intermediate pions are quite frequently bound in a Delta resonance and
in that intervals the real part of the pion optical potential is substituted
by the real part of the $\Delta$ optical potential.
Free intermediate  and final charged pions experience coulomb forces which
contribute to the real part of the pion optical potential.
Recent investigations on the influence
of the nuclear medium correction to the dispersion
relation of the free pion have shown conflicting
results \cite{ehe93,xio93} with respect to the importance of
the modification for low momentum pions. However, both
calculations show that the
high energy part of the pion spectrum remains unchanged by this
modification.  Since our results are mainly for this high
energy contribution we omit this  medium correction, until a consensus has
been achieved on the proper form of the respective medium contribution.

After a pion is produced (be it free or {\em bound} in a delta),
it's fate is governed by two distinct processes:
\begin{enumerate}
\item absorption
        $\quad \pi \, N \, N  \rightarrow  \Delta \, N
\rightarrow N \, N \qquad$
\item scattering (resorption)
        $\quad \pi \, N  \rightarrow  \Delta \rightarrow  \pi \, N $
\end{enumerate}
In the CASCADE mode all real forces are turned off:
nucleons, pions and deltas are
propagated on straight lines between  collisions.

\section{Formation and spectroscopy of $\Delta$ resonance matter}
Recently an old subtopic \cite{st78,cha73} of this
research has received renewed attention \cite{bog,wald,metag,ehehalt,mhof}:
The possibility of producing
{\em $\Delta$-matter} (or in more general terms: {\em resonance matter}). At
beam energies above a few hundred MeV/nucleon,
the nucleons can be excited into
$\Delta$-resonances. If the density of these resonances
is as high  as the nuclear
matter ground state density, then a new state of matter,
{\em $\Delta$-matter}, has
been created.
One of the potential signals for the presence of {\em $\Delta$-matter}
is the creation of pions as decay-products of the $\Delta$-resonance.

\begin{figure}
\centerline{\psfig{figure=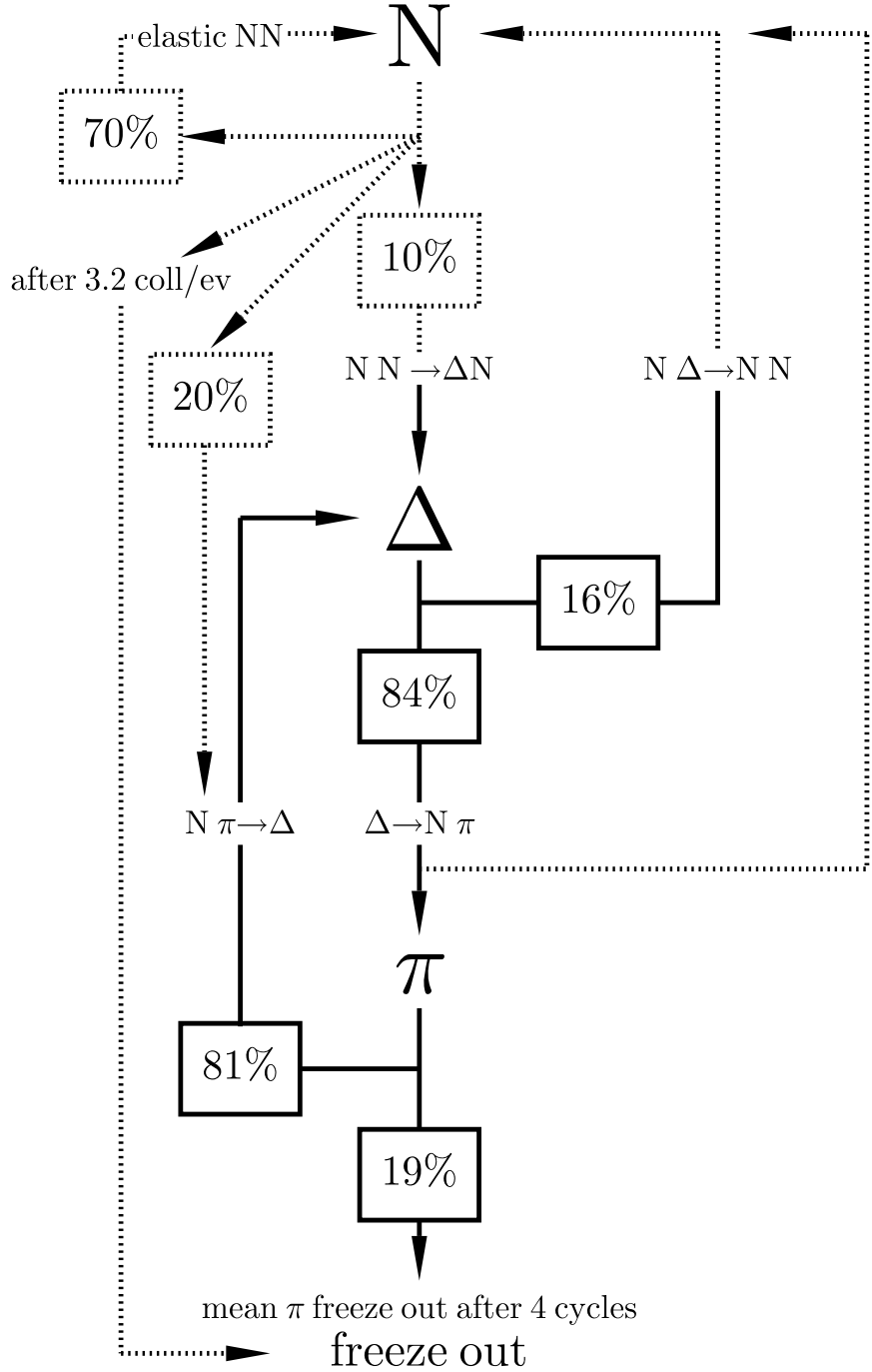,height=19cm}}
\caption{\label{pumpe} Pion - nucleon cycle in the IQMD model.
The scheme describes (for b
$\le 5$ fm and time-averaged) all in the model possible processes linked to
the
creation of {\em $\Delta$-matter}. The probabilities in the boxes always
refer t
o the
vertices they are directly connected with. The main process for sustaining
{\em $\Delta$-matter} is the $\Delta \to N \pi \to \Delta$ loop, which,
however,
first has to be fueled by the $N N \to \Delta N$ process.}
\end{figure}

How can {\em $\Delta$-matter} be produced? Figure \ref{pumpe} shows the
pion -- nucleon cycle in the IQMD model.
The scheme describes (for impact parameters b
$\le 5$ fm and averaged over 60 fm/c
possible processes linked to the
creation of {\em $\Delta$-matter}.
The probabilities in the boxes always refer to the
vertices they are directly connected with.
$\Delta$-resonances are initially produced via inelastic nucleon
nucleon scattering. The produced resonances can either be reabsorbed  via
inelastic scattering or decay by
emitting a pion. The pion can then either {\em freeze out}
or interact with a nucleon to  form a
$\Delta$ again. In case the $\Delta$ has been
absorbed the corresponding high energetic ergetic
nucleon might have a second chance of becoming a $\Delta$ by inelastic
scattering. It could also transfer energy via elastic scattering onto another
nucleon which then could scatter inelastically and form a new $\Delta$.
A nucleon interacts in the average about three times before it freezes out.
This value fluctuates considerably, depending on whether the
nucleon was in the participant zone (geometrical overlap of the colliding
heavy
ions) or in the spectator zone of the collision.

Unfortunately, the probablity for a nucleon to undergo inelastic
scattering and to form a $\Delta$ during the heavy ion collision is as low as
 10\%. The main process for sustaining
{\em $\Delta$-matter} is the $\Delta \to N \pi \to \Delta$ loop,
which, however,
first has to be fueled by the $N N \to \Delta N$ process.
The average pion passes approximately three times through this loop (it has
been created by the decay of a {\em hard} $\Delta$). However,
30\% pass more than 6 times through the loop.
For nucleons the probability of forming a {\em soft} $\Delta$ i.e. via
$\pi N \to \Delta$
is almost twice as high ({\em $\Delta$-matter pump})
than the probability of forming a {\em hard} $\Delta$
via $N N \to N \Delta$.

\begin{figure}
\centerline{\psfig{figure=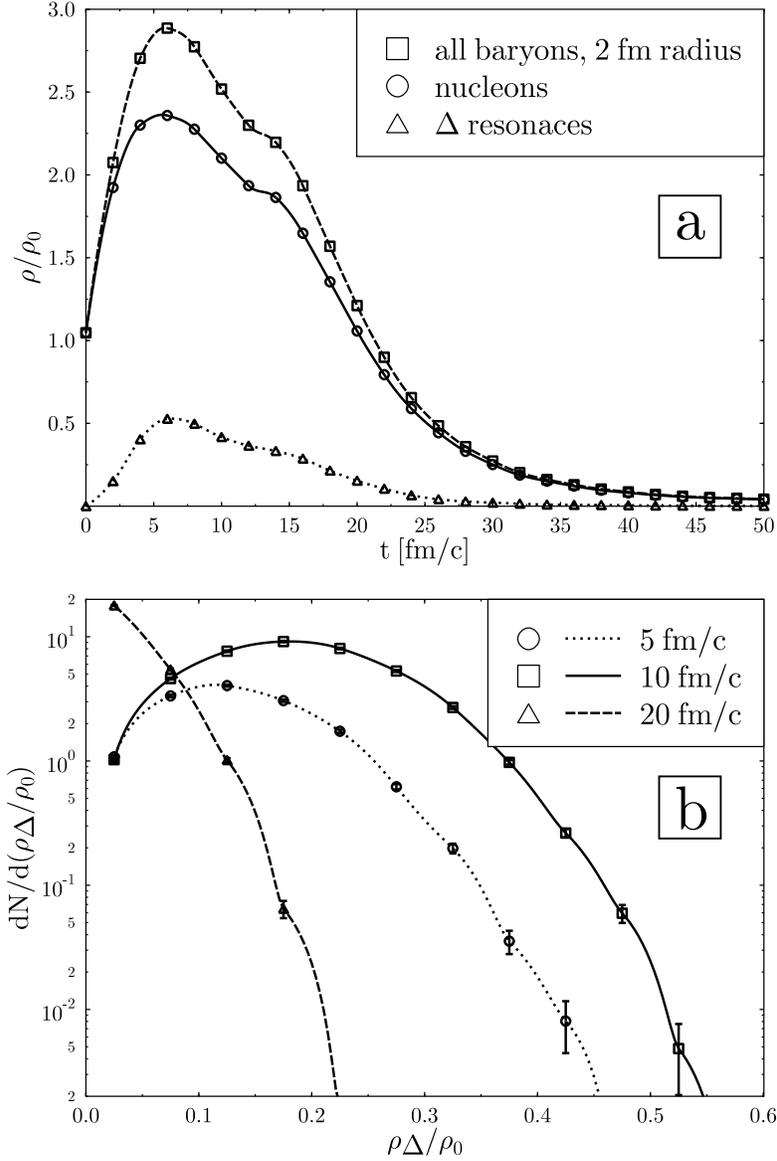,height=17cm}}
\caption{\label{tevol} Time evolution of the
the total baryon, nucleon and $\Delta$-resonance
density in units of $\rho/\rho_0$ (a) and $\Delta$ density distribution the
respective $\Delta$s experience for 5, 10 and 20 fm/c.
The densities in the upper frame (a) are calculated in a
sphere of 2 fm radius around the
collision center. The hot and dense reaction phase lies between 5 and 20 fm/c
during which approximately 10\% of the nucleons are excited to
$\Delta$-resonances.
Up to 50\% nucl. matter ground state density is reached by the
$\Delta$-resonances.
However, less than 1\% of the $\Delta$s experience such high densities.
The average
$\Delta$ density which is {\em felt} by the $\Delta$s is
approximately 0.25 $\rho_0$ at 10 fm/c.
The densities in the lower frame (b) were calculated by summing over all
contributing Gaussians of all
$\Delta$'s in the system at the locations of the respective $\Delta$'s.
}
\end{figure}

Figure \ref{tevol}a) shows the time evolution of
the total baryon, nucleon and $\Delta$ densities
in units of $\rho/\rho_0$ (top).
The densities are calculated in a sphere of 2 fm radius around the
collision center.
Between 5 fm/c and 20 fm/c
more than 20 $\Delta$-resonances can be found in the system:
This time interval coincides with the hot and dense reaction phase.
At 10 fm/c up to 55
resonances are present in the {\em total} reaction volume (keep in mind
this is {\em not} in the 2 fm test sphere).
A $\Delta$ multiplicity of $> 40$ can be sustained for an interval of 10 fm/c,
6 times longer than the lifetime of a free $\Delta$-resonance. However, this
is not pure {\em $\Delta$-matter}:
in the small {\em test} volume shown in figure 2a the
resonance {\em density} is 0.5 $\rho_0$ and the
nucleon density is 2.2 $\rho_0$: the $\Delta$-contribution is 20\% in the
test volume which contains, as a matter of fact, only 2.5 resonances.
The total multiplicity of $\Delta$ resonances  is just about
10\% of the total nucleon multiplicity.

However,
it is obvious that the other $\Delta$s can be distributed all over the
reaction volume.
Figure \ref{tevol}b) shows the $\Delta$ density
distribution as experienced by the
$\Delta$'s in the system at 5, 10 and 20 fm/c.
The densities were calculated by summing over all contributing Gaussians of
all
$\Delta$s in the system at the locations of the respective $\Delta$s.
We would like to point out that the mean $\Delta$-density experienced by
the $\Delta$s is about 0.25 $\rho_0$. Less than 1\% of the $\Delta$s
experience $\Delta$ densities around 0.5 $\rho_0$.
However, enough $\Delta$s are in the system to show signs of collectivity
such as {\em collective flow} in the reaction plane. Its measurable
signiture (the pion $\vec{p_x}(y)$ distribution in central collisions)
will be discussed in one of the follwing sections.

\section{Inclusive pion observables}
\begin{figure}
\centerline{\psfig{figure=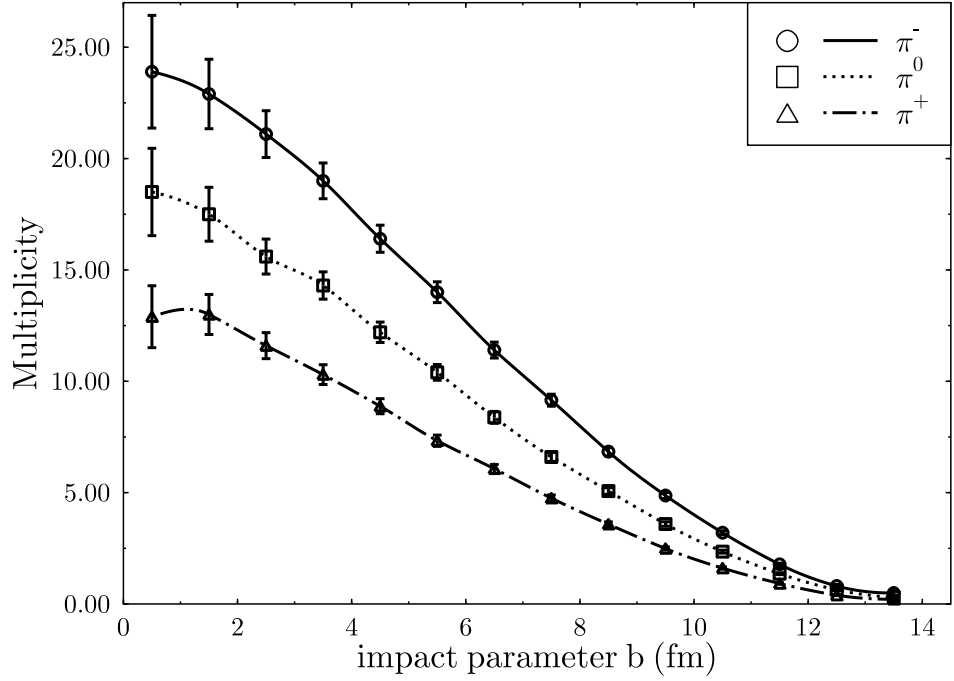}}
\caption{ \label{pimult} Multiplicity of $\pi^-, \pi^0$ and $\pi^+$
versus impact parameter $b$ for Au+Au collisions at 1 GeV/nucl. incident
beam energy. A hard EoS with momentum dependent interaction is used.
For central collisions ($b=0$ fm) the total pion multiplicity is
approximately 55. For a minimum bias impact parameter distribution the average
pion multiplicity is about 19 (8 $\pi^-$, 6 $\pi^0$ and 5 $\pi^+$). }
\end{figure}

This section deals with inclusive pion production in
relativistic heavy ion collisions.
Figure \ref{pimult} shows the
predicted impact parameter dependence of the
multiplicity of $\pi^-, \pi^0$ and $\pi^+$
for Au+Au collisions at 1 GeV/nucl. incident
beam energy.
A hard EoS with momentum dependent interaction is used in the calculation.
For central collisions ($b=0$ fm) the total pion multiplicity is
approximately 55. However,  the average
pion multiplicity is about 19 (8 $\pi^-$, 6 $\pi^0$ and 5 $\pi^+$)
for a minimum bias impact parameter distribution.

\begin{figure}
\centerline{\psfig{figure=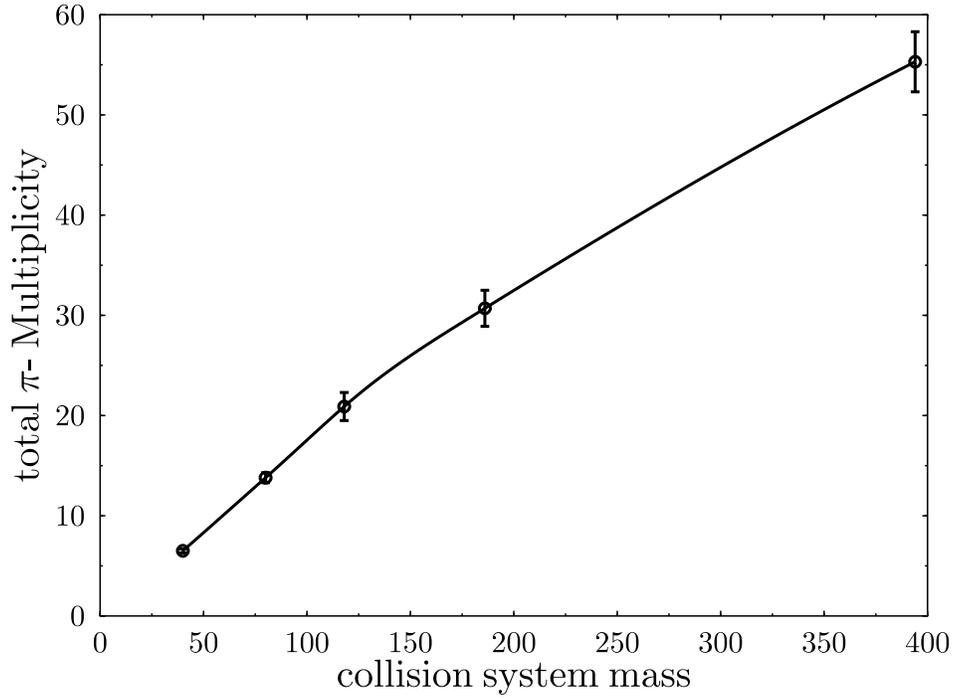}}
\caption{ \label{pimult-a} Total pion multiplicity versus collision system
mass at 1 GeV/nucl. beam energy.
For light collision systems the multiplicity increases linearly with
the system mass. However, for heavy systems the increase is less than linear
due to pion absorption. The values plotted were extracted from $b=0$ fm
calculations of the systems Ne+Ne, Ca+Ca, Ni+Ni, Nb+Nb and Au+Au. }
\end{figure}

The mass dependence of the total pion multiplicity is shown in
figure \ref{pimult-a}.
For light collision systems the multiplicity increases linearly with
the system mass. However, for heavier systems the increase is less than the
 linear extrapolation, this is
due to pion absorption. The values are for $b=0$ fm
calculations of the systems $^{20}$Ne+$^{20}$Ne, $^{40}$Ca+$^{40}$Ca,
$^{58}$Ni+$^{58}$Ni, $^{93}$Nb+$^{93}$Nb and $^{197}$Au+$^{197}$Au.

\begin{figure}
\centerline{\psfig{figure=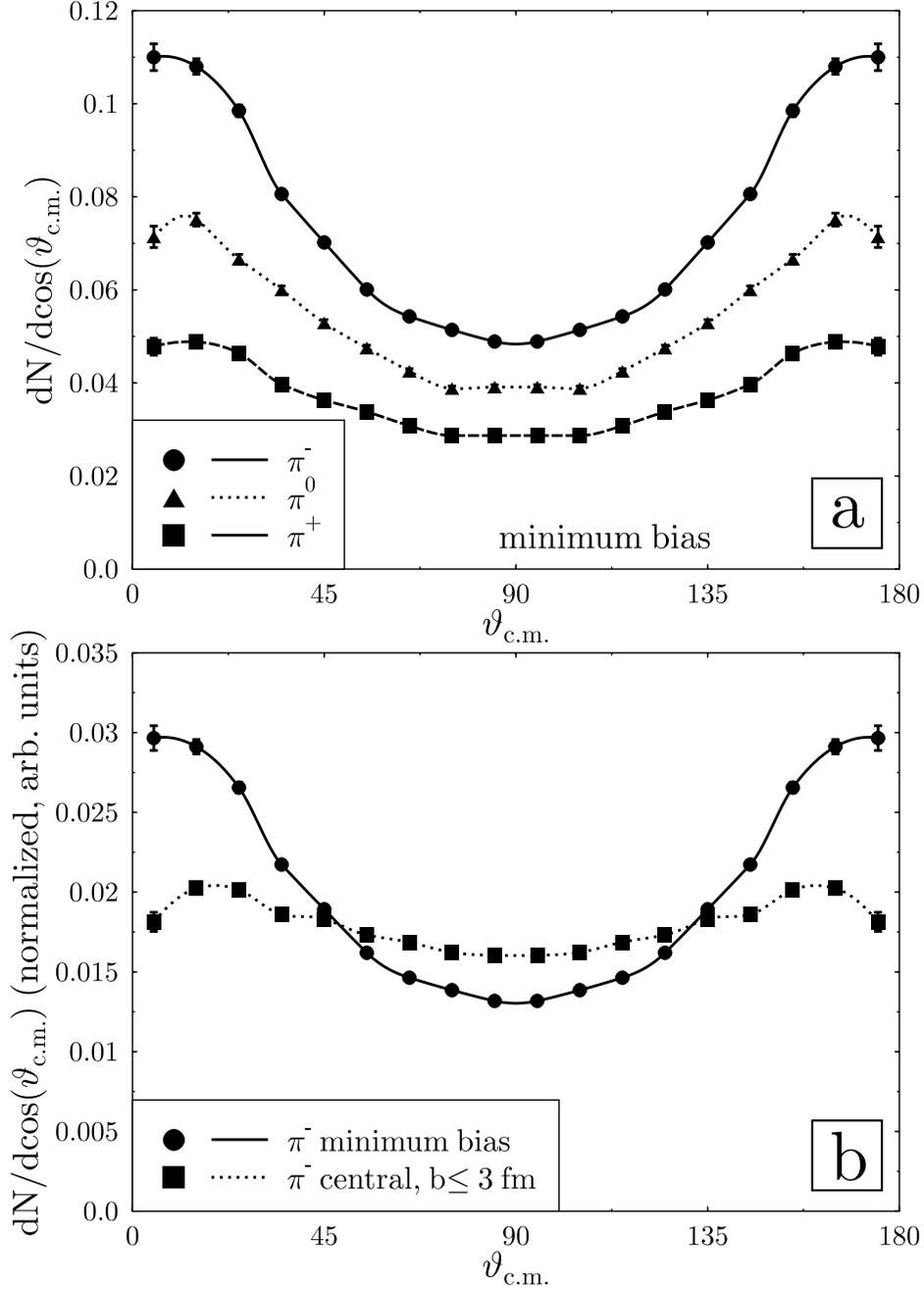,height=19cm}}
\caption{\label{pi-theta} Polar angular distribution $\frac{dN}
{d\cos\vartheta_{c.m.}}$
for $\pi^-,\pi^0$ and $\pi^+$ in minimum bias  (a)  and for $pi^-$ in
minimum bias and central (b) Au+Au collisions. A horizontally flat
distribution would
correspond to an isotropic emission. For minimum bias events (top) a strong
peaking
towards forward--backward angles is observed, most prominently for $\pi^-$.
The anisotropy decreases when studying central collisions (bottom). It can be
explained by the decay of $\Delta$-resonances in the projectile- and target-
 spectator regions. }
\end{figure}

The polar angular distribution $\frac{dN}{d\cos\vartheta_{c.m.}}$
for $\pi^-,\pi^0$ and $\pi^+$ is shown in Figure \ref{pi-theta}
for minimum bias events (a) and for $\pi^-$
in central vs. minimum bias  events in (b).
A horizontally flat distribution would
correspond to isotropic emission.
For minimum bias events (top) a strong peaking
towards forward--backward angles is observed, most prominently for $\pi^-$.

This is important for the extrapolation of total yields from spectra
measured at $\vartheta_{c.m.}= 90^{\circ}$ -- if the midrapidity spectra are
used to extrapolate (with the assumption of a flat distribution)
the total yield may be underestimated by a factor of 2.
The anisotropy decreases when studying central collisions (bottom).
This dependence can be
explained by the decay of $\Delta$-resonances in the projectile- and target-
spectator regions.
The difference betweeen the distributions of $\pi^-$ and $\pi^+$ also results
in a forward--backward peaking of the $N_{\pi^-}/N_{\pi^+}$ ratio.
This phenomenon
has already been experimentally observed for light collision systems
at the BEVALAC \cite{fuji}.

Figure \ref{tapsspek} shows a comparison of inclusive $\pi^0$ spectra
for Au+Au and Ca+Ca (minimum bias and $y_{c.m.}\pm 0.16$) between the IQMD
model and data published by the TAPS collaboration \cite{taps2}. Whereas
the model shows reasonable agreement with
the heavy system Au+Au it overpredicts
the $\pi^0$ yield of the light system Ca+Ca by approximately 60\%.
The charged pions (for Au+Au collisions) are shown in
figure \ref{kaosspek} together with $\pi^+$ data from the KaoS
collaboration \cite{chm94}.
The slope of the $\pi^+$ spectrum in
the model calculation agrees well with the KaoS measurements.
However, the multiplicity as predicted by the model
is approximately 20\% above the KaoS measurements.
Both, the calculation and the measurements, have been acceptance corrected
to the rapidity interval $y_{c.m.}\pm 0.16$ and may directly be compared
to Figure \ref{tapsspek}. Especially the measurement of high energy
pions is of great interest. They correlate directly to early freeze out
times and heavy $\Delta$ resonances \cite{ba94a}.

The mass dependence of pion production and its sensitivity towards the
transverse momentum $p_t$ can be studied more clearly by plotting the ratio
of the $p_t$ spectra for Au+Au and Ca+Ca versus the transverse momentum $p_t$
(Figure \ref{masspt}). A comparison between data from the TAPS collaboration
\cite{taps2} and the IQMD model is shown in figure \ref{masspt}.
The model underpredicts this ratio by approximately
a factor of 2, In particular for low transverse momenta .
However, this holds for all transport model calculations which have been
compared to the TAPS data \cite{taps2}. A previous comparison between the
IQMD model and the TAPS data in ref. \cite{taps2}
showed far larger disagreement between model
and data. This was due to an improper
normalization of  the theoretical calculations.

The yield of low $p_t$ pions in the heavy system is underpredicted by
10\%. This is a common problem of most transport theories dealing with
heavy ion collisions. Suggested explanations for this underprediction include
in-medium effects of pions in nuclear matter \cite{xio93} and
the neglect of Bose-enhancement
due to the bosonic nature of the pions.

The importance of the inclusion of Coulomb forces and energy dependent
$\pi - N$ cross sections can be shown by plotting the $\pi^-$ to $\pi^+$
ratio versus the transverse momentum $p_t$ (Figure \ref{picoulomb}).
The solid line
shows the full calculation including Coulomb forces. For high $p_t$ the
$\pi^-/\pi^+$ ratio decreases towards 1, whereas for low $p_t$
it increases to 2.5 --
considerably higher than the value of 1.8 predicted by the $\Delta$-isobar
model.
The dashed line shows a calculation  without Coulomb forces.
This ratio remains constant at 1.8.
The (small) remaining variations might be
due to the different energy dependence of the $\pi^+ - p$ and $\pi^- - p$
inelastic cross sections.

\section{ Pion nucleon correlation in the event-plane}
The hydrodynamical model predicts a bounce--off of nuclear matter
in the reaction plane \cite{st80,st79} which has experimentally indeed been
discovered \cite{cse82,gut79}. The  bounce--off is depicted by plotting
the in--plane transverse momentum $\vec{p_x}(y)$ versus the
rapidity $y$.  For nucleons and light fragments a horizontal s--shape is
typically seen with negative $\vec{p_x}(y)$ values for
$y \le y_{c.m.}$ and positive $\vec{p_x}(y)$ values for
$y \ge y_{c.m.}$.
Figure \ref{piflow} shows the
$\vec{p_x} (y)$ distribution for $\pi^+$ and protons in
Au(1AGeV)Au collisions with a minimum bias impact parameter distribution.
The protons show the expected collective flow \cite{moli85,gus84a,moli87}.
The $\vec{p_x}$ of the pions, however, is anticorrelated to that of the
protons. A similar proton -- $\pi$ anticcorelation has been measured
for the asymmetric system Ne+Pb at 800 MeV/nucl.
by the DIOGENE collaboration \cite{gos89}.
Transport model comparisons to the DIOGENE data with the IQMD
\cite{ha88} and the BUU \cite{lib91a} model have shown good agreement with
the data.

We have studied the origin of the particular shape of the pion
angular distribution and the $\vec{p_x}$ spectrum
by sequentially
suppressing first the {\em soft-delta}-production and then the
delta-absorption (while allowing the {\em soft-delta}-production).
If we deactivate the {\em soft--delta}--production (see Figure 3),
$\pi \, N \rightarrow \Delta$,
pions are neither scattered nor absorbed after the initial production.
No $\vec{p_x}$ for pions is observed.
In order to decide whether the $\vec{p_x}$ spectrum is
caused by absorption or by scattering we now
deactivate the reaction
$\Delta \, N \rightarrow  N \, N $.
We thus suppress  pion absorption
but allow scattering -- the anticorrelation between pions and protons in the
$\vec{p_x}$ returns.
In contrast to previous publications, which
investigated the asymmetric system Ne(800AMeV)Pb and suggested the
anticorrelation
of pionic and nucleonic $\vec{p_x}$ at target rapidities to be
caused by pion absorption \cite{lib91a},
our investigation reveals the $\vec{p_x}$ spectrum of the pions
to be dominated by the pion scattering process \cite{baplb}.

The following simplified picture can explain the origin of the
observed phase space distribution:
The $\Delta$ decays isotropically in its rest--frame, therefore
50 \% of the pions are emitted with a positive $p_x$ and 50 \% with
a negative $p_x$.
At target rapidity those pions which obtain
a positive $p_x$--value usually do not have
the chance  to rescatter: Most of the target nucleons are
located in the {\em negative } $p_x $ {\em area!}
Those pions which {\em do} rescatter at target rapidity are the ones with an
initially {\em negative}
$p_x $: Every time a $\Delta$ decays (isotropically)
there is a 50\% chance that this pion is emitted {\em upward}, i.e. into an
azimuthal angle between
$-90^{\circ} \le \phi \le 90^{\circ} $. These $\phi$--values characterize
the hemisphere of
positive $ p_x $, by definition.
This leads
for$ \approx 50\%$ of the pions with
 -- originally -- negative $ p_x $ to a shift towards a positive $p_x$.
This remains true even after transforming back into the laboratory
frame. The same consideration applies vice versa for projectile rapidity:
Most projectile nucleons are located in the {\em positive} $p_x$ {\em area}.
The pions are rescattered in this area which results in a
negative $\vec{p_x }$
and a maximum in the azimuthal angular distribution in the
$90^{\circ} \le \phi \le 270^{\circ} $  interval.

Figure \ref{pipxeos} shows the in--plane transverse momentum $\vec{p_x} $
versus rapidity $y$ (in the c.m. system) for $\pi^+$ in central collisions
of Au+Au (with impact parameters b$\le 3$ fm).
In contrast to (semi-) peripheral collisions, however,
$\vec{p_x}$ is correlated for pions and
nucleons in central collisions because of the
bounce--off of $\Delta$-resonances \cite{ba94b}.
The square markers in Figure \ref{pipxeos} depict a calculation with the
hard equation of state without momentum dependence, the circles show
the same equation of state including momentum dependence whereas the
triangles represent a CASCADE calculation, i.e. a non-equilibrium
free gas.

The momentum dependence enhances the
$ \vec{p_x} $ of the pions. This effect is due to the bounce--off
of the $\Delta$ resonances \cite{ba94b} which in our model is  enhanced
because the  momentum dependence for the $\Delta$ resonances
is included  in the same way as for the nucleons.

The  CASCADE calculation, however, shows the opposite behaviour. The
$\vec{ p_x}$  of the pions has negative sign to that of the
calculations with the density dependent equations of state.
This behaviour can be explained by the
lack of hadron collective flow in CASCADE calculations \cite{moli86}.
The pions would then be expected to be emitted isotropically
($\vec{p_x}(Y)=0$).
However, pion scattering from small caps of spectator
matter being present at impact parameters around 3 fm causes the observed
anticorrelation \cite{baplb}.
In order to investigate the density dependence of the nuclear equation of
state
and in order to show the differences between CASCADE calculations
and calculations
including the equation of state more clearly we
use the robust observable $p_x^{dir}$ which for nucleons
is defined as
\begin{displaymath}
p_x^{dir} \,=\, \frac{ \sum\limits_{i=1}^{A_P+A_T} p_x^i
\cdot {\rm sgn}(y_i - y_{c.m.})}{A_T+A_P} \quad.
\end{displaymath}
(the adaptation for pions is straightforward)
and plot it versus the impact parameter (Figure \ref{pipxdir}).
For positive values of $p_x^{dir}$ the pion $\vec{ p_x }$ vs. rapidity
distribution is correlated to that of the nucleons. For negative values an
anticorrelation is observed.

Figure \ref{pipxdir}
shows the respective calculations for the hard and soft equations
 of state
(including momentum dependence)
and for the CASCADE calculation. For small impact parameters the
calculations with
equation of state show a correlation between pion and nucleon
bounce--off. At
semiperipheral impact parameters we observe a sign reversal.
As mentioned above,
the anticorrelation between nucleon and pion bounce--off is
caused by pion scattering in spectator matter \cite{baplb}.
In contrast, the CASCADE calculation exhibits a negative $p_x^{dir}$
for the whole impact parameter range.
The momentum transfer $p_x^{dir}$ shows a systematic difference between the
hard and soft equation of state.
However, very high statistics and high precision impact parameter
classification are  necessary to experimentally exploit this sensitivity
towards the determination of the nuclear equation of state.
The results of figures \ref{piflow} and \ref{pipxdir} show
clearly that even in the domain of particle production
($\pi, K, \eta, \bar{p}, \rho, \omega$) CASCADE simulations
predict distinctly different phase space
distributions for baryons and mesons at central impact parameters.

\section{Azimuthal correlations of pions pependicular to the event plane}
Now let us investigate particle emission perpendicular to the reaction plane.
The hydrodynamical model predicted a squeeze--out of high energetic
nucleons perpendicular to the reaction plane \cite{st82,st86,bu83}.
This effect, which
has also been predicted by QMD--calculations \cite{ha88,ha90,ch92,ch93,ba94c}
and has been confirmed by experiment \cite{gut89,gut89b,leifels},
is due to the high compression of nuclear matter in the central hot and
dense reaction zone (it is a genuinly collective effect, increasing
linearly with $A$).

Do pions show a similar behaviour?
The azimuthal ($\varphi$)
distribution of the pions is plotted to investigate this question.
$\varphi$ is the angle between
the transverse momentum vector $\vec{p_t}$ and the $x$-axis (which lies
in the reaction plane and is perpendicular to the beam axis).
Thus $\varphi=0^{\circ}$  denotes the projectile hemisphere
and $\varphi=180^{\circ}$  corresponds
to the target hemisphere.

Figure \ref{piphi} shows the respective distributions
for neutral pions in the transverse momentum bins
$p_t \le$ 50 MeV and $p_t \ge$ 400 MeV at a minimum bias
impact parameter distribution.
The distributions have been normalized in order
to fit into the same figure.
The analysis was performed from $0^{\circ}$ to $180^{\circ}$ and then
symmetrized for $180^{\circ}$ to $360^{\circ}$.
The plotted distributions have been extracted by fitting
the calculated points (shown for the high $p_t$ bin)
according to the function $a\cdot (1+ b\cdot \cos(\phi) + c\cdot
\cos(2\phi))$.
The azimuthal angular distribution for $\pi^0$ with low $p_t$ shows maxima
at $\varphi=0^{\circ}$ and $\varphi=180^{\circ}$ corresponding to a
preferential emission in the reaction plane.
The high $p_t$ $\pi^0$, however, show
a maximum at
$\varphi=90^{\circ}$.
This maximum is associated with preferential particle emission
perpendicular to the reaction plane.
The inlay shows data from the TAPS collaboration \cite{ven93} for
the region 400 MeV $\le p_t \le$ 600 MeV and midrapidity. We observe a good
qualitative agreement between the theoretical prediction and the experiment.
It should be noted, however, that both, theory and experiment, need much
better statistics to allow a conclusive quantitative comparison.

The magnitude of the observed anisotropy and its dependence
on impact parameter and transverse momentum is
best studied by using the following ratio:
\begin{displaymath}
R_{out/in} \,=\, \left.\frac{ \frac{dN}{d\varphi}(\varphi=90^{\circ}) +
\frac{dN
}{d\varphi}(\varphi=270^{\circ})}
                            { \frac{dN}{d\varphi}(\varphi=0^{\circ}) +
\frac{dN}
{d\varphi}(\varphi=180^{\circ})} \right|_{y=y_{c.m.}}
\end{displaymath}
For positive $R_{out/in}$ values pions are emitted preferentially
perpendicular
to the reaction plane.
Figure \ref{pisqr} shows the transverse momentum dependence of $R_{out/in}$
for
Au+Au collisions with an impact parameters bewteen b=5 fm and b=10 fm and
at midrapidity ($y_{c.m.} = \pm 0.2$):
In contrast to pions with low transverse momentum, which are emitted
preferentially in the reaction plane,
high $p_t$ pions are preferentially emitted
perpendicular to the reaction plane. This effect is stronger for
$\pi^+$ than for $\pi^-$. The difference is due to
the different $\pi N \to \Delta$ production cross section for
$\pi^+$ and $\pi^-$ and due to Coulomb forces pushing the $\pi^+$ away from
the spectator matter which is located mostly in the reaction plane. The
$\pi^-$ on the other hand are being attracted by those spectator-protons.
These effects decrease the number of $\pi^-$ leaving the reaction zone in a
direction perpendicular to the reaction-plane. However, the statistics
accumulated
so far are not large enough for a more detailed study of the differences
between
positive and negative pions.
The inlay of Figure \ref{pisqr} shows
recent measurements from the KaoS collaboration \cite{brill93a} which confirm
the
predicted systematics of the $p_t$ dependence.
Imposing the limited acceptance of the
KaoS spectrometer on the IQMD calculations would reduce the available
statistics by one order of magnitude. Experimental uncertainties
in the determination of the proper reaction plane result in a reduction of the
measured $R_{out/in}$ values which are difficult to compensate. Therefore
a direct quantitative comparison between the KaoS measurements and IQMD
calculations is not feasable at this point of time.

We have investigated the cause of the observed preferential emission
perpendicular
to the reaction plane:
Pion absorption as well as scattering can be eliminated
by deactivating the reaction
$\pi \, N \rightarrow \Delta$, then no sqeeze-out is observed.

In order to decide whether the anisotropy is
caused by absorption or by scattering
the reaction
$\Delta \, N \rightarrow N \, N $ can be deactivated.
Thus pion absorption is suppressed
but scattering is allowed: no anisotropy
is observed. Therefore we conclude that the anisotropy is dominated
by the pion absorption process \cite{baprl}.

Figure \ref{pindel} shows the distribution of the
number of delta generations $n_{\Delta}$
a pion goes through before its freeze out.
Here $n_{\Delta}$ is shown for $\pi^+$ emitted both
in the reaction plane as well as perpendicular to it.
$(n_{\Delta}-1)$ is therefore the number of times
a pion scatters before freeze out. We observe that 90\% of the produced
pions scatter at least once before leaving the reaction zone. A large
number of pions scatters even more often, 2\% up to 10 times! The
observed preferential emission perpendicular to the reaction plane is due
to an excess of high $p_t$ pions which
on the average have undergone fewer collisions ($\le2$) than the pions
in plane.
Those pions which make this effect do rescatter rarely, they are emitted
early but carry information on the high density phase of the reaction.
They stem from the decay of the most massive delta resonances
which are mostly produced early on in the reaction \cite{ba94a}.
Therefore high $p_t$ pions emitted perpendicular to the
event-plane  should be the most sensitive pionic probes for the investigation
of the hot and early reaction zone.

Figure \ref{piphieos1} shows the azimuthal angular distribution of high $p_t$
($p_t \ge 400$ MeV) neutral pions at midrapidity and impact parameter
b=6 fm. The different curves show
calculations for hard (circles) and soft (squares)
equations of state (including
momentum dependence) and a CASCADE calculation (triangles).
$\varphi$ is the angle between
the transverse momentum vector $\vec{p_t}$ and the $x$-axis (which lies
in the reaction plane and is perpendicular to the beam axis).
The out--of--plane pion squeeze--out is clearly seen by the pronounced
maximum at $\varphi = 90^{\circ}$ for both equations of state.
To enhance the statistics all
particles are projected into the $0^{\circ} \le \varphi \le
180^{\circ}$ hemisphere.
The full and dashed lines are least square fits with the function
$f(\varphi)= a ( 1 + s_1 \cos(\varphi) + s_2 \cos(2 \varphi))$ which has
been  used to fit the squeeze--out phenomenon \cite{gut89,gut89b}.
The curves show an
extrapolation to the full azimuthal angular range.
The distributions are normalized per particle
in order subtract the influence of
different equations of state on the pion multiplicity.
Within error-bars both equations of state exhibit the same
out--of--plane pion squeeze--out.
There is a trend for the hard equation of state to exhibit an enhanced
out--of--plane pion squeeze--out but this trend might be too small to be
useful for an experimental distinction between the different equations of
state, whereas -- in contrast -- the in--plane pion bounce--off shows
a clear difference for the two cases (see above).
The CASCADE calculation
does not exhibit any significant out--of--plane pion squeeze--out
for b=6 fm!
However, for larger impact
parameters also CASCADE calculations exhibits a
pronounced out--of--plane pion squeeze--out
(Figure \ref{pisqb}).

Figure \ref{pisqb} shows $R_{out/in}$ versus impact parameter b for the hard
equation
of state with and without momentum dependence of the real part of the
nucleon optical potential. The momentum
dependence causes a drastic increase of $R_{out/in}$ for impact
parameters larger than 3 fm. In the CASCADE calculation the onset of
the out--of--plane pion squeeze--out is shifted toward larger
impact parameters as compared
to the calculations including the equation of state (see also
Figure \ref{piphieos1}).
For peripheral collisions $R_{out/in}$ reaches the same magnitude for
the CASCADE calculation and the hard equation of state without momentum
dependence.

\section{Conclusion and outlook}
The physics responsible for the
in--plane {\em pion} bounce--off (pion scattering) and
the out--of--plane {\em pion} squeeze--out (pion absorpion)
differs completely from
the compressional effects governing the in--plane {\em nucleon}
bounce--off and out--of--plane {\em nucleon}
squeeze--out.
It is the pion--nucleon interaction which creates the sensitivity towards
the density and momentum dependence
of the nucleon optical potential. Therefore it is understandable that
we observe
a strong sensitivity towards the equation of state {\em in}
the reaction plane whereas
the sensitivity towards the equation of state
{\em perpendicular} to the reaction plane
is limited to the momentum dependence:
The (anti-) correlation in--plane is caused by multiple
pion nucleon scattering \cite{baplb} with the bounced--off nucleons,
which show a strong sensitvity towards momentum and density dependence
\cite{ha89}.
The pion squeeze--out perpendicular to the reaction plane, however,
is dominated by high $p_t$ pions which have undergone
less rescattering than those in the reaction plane \cite{baprl}. The abundance
of these high $p_t$ pions is correlated to the multiplicity of high $p_t$
nucleons which increases if the density dependence is included.

We have investigated the dependence of {\em pionic} in--plane
bounce--off and out--of--plane squeeze--out
on the {\em nuclear} equation of state. A strong sensitivity towards the
{\em density dependence} is observed for
the in--plane {\em pion} bounce--off whereas the
out--of--plane {\em pion} squeeze--out shows only a small sensitivity.
Both effects show a strong sensitivity toward the {\em momentum dependence}.
CASCADE caluculations -- which we see as a crude approximation to QMD --
give different phase space
distributions for pions in both cases.
It should be easy to resolve experimentally these two clearly qualitatively
different distinct scenarios. The determination of the equation of state
will require -- on the other hand -- a more sensitive (and sensible)
quantitative comparison to theory, including a improved
treatment of the $\Delta$ and pion optical potential.

The nuclear equation of state cannot be
extracted from one observable alone. All observables known to be sensitive
to the equation of state have to be fitted simultaneously by the respective
model in order to claim success. In this article we have added additional,
here {\em pionic},
observables which have to be taken into account for obtaining the final
goal: The nuclear equation of state.

\section{Acknowledgements}
This work was supported by GSI, BMFT and DFG.
Most calculations presented in this article were performed
at the computing center of the
University of Frankfurt  on a
Siemens-Nixdorf SNI 200-10 (Fujitsu VP)
super-computer (comparable to a Cray Y-MP) and  required approximately
700 cpu-hours.
Additional calculations (approximately 250 cpu hours)
and the entire analysis were
performed on the IBM RISC 6000 model 580/590 cluster
of the GSI computing department. We wish to thank both institutions for their
continuing support of our research efforts.

Furthermore we wish to thank numerous members of the KaoS, TAPS, FOPI and EOS
collaborations, especially
C. M\"untz, H. Oeschler, R. Simon, V. Metag, H. G. Ritter, D. Brill,
L. Venema, H. L\"ohner,  M. Trzaska, P. Senger, E. Grosse , Th. Wienold,
N. Herrmann and C. Pinkenburg for many fruitful discussions and their support
of our work.


\pagebreak

\begin{table}
\begin{tabular}{lccccc}
 &$\alpha$ (MeV)  &$\beta$ (MeV) & $\gamma$ & $\delta$ (MeV) &$\varepsilon \,
 \left(\frac{c^2}{\mbox{GeV}^2}\right) \!\!\!\!$ \\
\hline
 S  & -356 & 303 & 1.17 & ---  & ---    \\
 SM & -390 & 320 & 1.14 & 1.57 & 500  \\
 H  & -124 & 71  & 2.00 & ---  & ---    \\
 HM & -130 & 59  & 2.09 & 1.57 & 500  \\
\end{tabular}
\caption{\label{eostab} Parametersets for the nuclear equation of state used
in the
IQMD model. S and H refer to  the soft and hard equations of state, M refers
to the
inclusion of momentum dependent interaction.}
\end{table}

\begin{table}
\begin{tabular}{ccc}
$x=\sqrt{s}$ (GeV) & $a$ (fm) & $b$ \\\hline
2.104 -- 2.12 	&  $294.6 \; (x - 2.014)^{2.578}$
	& $19.71\; (x - 2.014)^{1.551}$   \\
2.12 -- 2.43	& $\frac{0.01224}{(x- 2.225)^2 + 0.004112}$
	& $19.71\; (x - 2.014)^{1.551}$ \\
2.43 -- 4.50 	& $(2.343/x)^{43.17}$ & $33.41 \; \arctan( 0.5404
\;(x-2.146)^{0.9784})$  \\
\end{tabular}
\caption{ \label{inelast} $a(s)$ and $b(s)$ as functions of the c.m. energy. }
\end{table}

\pagebreak

\begin{figure}
\caption{ \label{tapsspek} Comparison of inclusive $\pi^0$ spectra
$\frac{d\sigma}{p_t dp_t}$ for Au+Au and Ca+Ca (minimum bias) collisions
between the IQMD model and data measured by the TAPS collaboration.
A hard EoS including momentum dependence is used and  a rapidity cut
according to the acceptance of the TAPS spectrometer is employed.
The model shows reasonable agreement with
the heavy system Au+Au but it overpredicts
the $\pi^0$ yield of the light system Ca+Ca by approximately 60\%.
The yield of low $p_t$ pions in the heavy system is underpredicted by
10\%.}
\end{figure}

\begin{figure}
\caption{ \label{kaosspek} Inclusive $\pi^-$ and $\pi^+$ spectra
$\frac{d\sigma}{p_t dp_t}$ for Au+Au (minimum bias) collisions at
1 GeV/nucl. as calculated with the IQMD model and a $\pi^+$ spectrum
measured by the KaoS collaboration.
A hard EoS including momentum dependence is used in the IQMD
calculation and  a rapidity cut
according to the acceptance of the TAPS spectrometer is employed; the KaoS
measurement has been acceptance-corrected.
The slope of the $\pi^+$ spectrum in
the model calculation agrees well with the KaoS measurements.
However, the multiplicity as predicted by the model
is approximately 20\% above the KaoS measurements.
}
\end{figure}

\begin{figure}
\caption{ \label{masspt} Ratio of the pion yield from Au+Au and Ca+Ca
collisions plotted versus the transverse momentum $p_t$. The figure
shows a comparison between data measured by the TAPS collaboration and
an IQMD calculation (minimum bias, hard equation of state with momentum
dependence). For low transvere momenta the model underpredicts the data
approximately by a factor of 2.}
\end{figure}

\begin{figure}
\caption{ \label{picoulomb}$\pi^-$ to $\pi^+$
ratio versus transverse momentum $p_t$ for Au+Au
collisions (minimum bias) at 1 GeV/nucl. with a hard EoS and momentum
dependence. The solid line
shows the full calculation including Coulomb forces. For high $p_t$ the
ratio decreases towards 1 whereas for low $p_t$ it increases to 2.5 --
considerably higher than the value of 1.8 which the $\Delta$-isobar model
would suggest. The dashed line shows a calculation  without Coulomb forces
acting upon the pions.
Within the errorbars the ratio remains constant at a value around 1.8. }
\end{figure}


\begin{figure}
\caption{ \label{piflow}  Rapidity y vs. $\vec{p_x}/ m $
for $\pi^+$ and protons in
Au+Au collisions at 1 GeV/nucl. with minimum bias impact parameter
distribution
and a hard EoS including momentum dependence.
The protons show the expected bounce-off. The
$\vec{p_x}$ of the pions is directed oppositely to that of the protons.
This effect is caused by rescattering of pions from large chunks of
spectator matter.}
\end{figure}

\begin{figure}
\caption{ \label{pipxeos}
in--plane transverse momentum $\vec{ p_x} $
versus rapidity $y$ (in the c.m. system) for $\pi^+$ in central Au+Au
collisions with
impact parameters b$\le 3$ fm. Calculations with a hard equation of state
without momentum dependence (squares),  the same equation of state
with momentum dependence  (circles) and a CASCADE calculation
(triangles) are shown. The effect of the momentum dependence  is
considerable, exhibiting the sensitivity of $\vec{p_x}$ on the
baryon flow.
The CASCADE calculation gives a different phase space distribution due to its
lack of collective baryon flow. }
\end{figure}

\begin{figure}
\caption{ \label{pipxdir} $p_x^{dir}$ versus impact parameter $b$
for Au+Au collisions at 1 GeV/nucl. with hard and soft equations
of state, both with momentum dependence,
and for the CASCADE calculation. Note the clear sensitivity for
the equation of state on $p_x^{dir}$. The CASCADE calculation  exhibits
an anticorrelation between pions and nucleons for the whole impact parameter
range, due to pion nucleon scattering and its lack of collective baryon flow.}
\end{figure}


\begin{figure}
\caption{ \label{piphi}
Normalized azimuthal angular distribution $dN/d\varphi$ for $\pi^0$ with
low and high transverse momentum $p_t$ at mid-rapidity
in the reaction Au(1AGeV)Au with  minimum bias impact parameter
distribution, a hard equation of state and momentum dependent interaction.
The points were fitted according to the function
$a\cdot (1+ b\cdot \cos(\phi) + c\cdot \cos(2\phi))$.
The maximum at $\varphi=90^{\circ}$
corresponds to a preferential emission of high $p_t$ pions perpendicular
to the reaction plane. This is due to pion absorption
by large pieces of baryonic spectator matter located predominantely
in the reaction plane.
Perpendicular to the plane there is no such spectator matter and pions
with high $p_t$ can leave the reaction zone without
further interaction. Low $p_t$ pions have rescattered more often which
is only possible in the reaction plane.
The inlay shows data from the TAPS collaboration for
the region 400 MeV $\le p_t \le$ 600 MeV and midrapidity. }
\end{figure}

\begin{figure}
\caption{ \label{pisqr}Squeeze-out ratio $R_{out/in}$ versus transverse
momentum $p_t$ for $\pi^+$ and $\pi^-$.
Pions with $p_t\ge 200$ MeV are preferentially emitted perpendicular
to the reaction plane. Pions with $p_t \le 100$ MeV are
emitted isotropically because they have undergone
frequent rescattering which can only happen
due to spectator matter in the reaction plane. The differences between
$\pi^-$ and $\pi^+$ are due to Coulomb forces.
The inlay shows data on $\pi^+$ from the KaoS collaboration.}
\end{figure}

\begin{figure}
\caption{ \label{pindel}Distribution of the number of
delta generations $n_{\Delta}$ a pion goes through before its freeze out
for $\pi^+$ emitted in the reaction plane
and perpendicular to it.
90\% of the produced
pions scatter at least once before leaving the reaction zone.
The observed preferential emission
perpendicular to the reaction plane is due
to an excess of pions which
on the average have undergone fewer collisions ($\le2$) than the pions
in plane. }
\end{figure}

\begin{figure}
\caption{ \label{piphieos1}Azimuthal angular distribution $dN/d\varphi$
for neutral pions calculated
with hard and soft equations of state (both with momentum dependence)
and a CASCADE calculation.
All calculations were performed with an impact parameter
of b=6 fm. Both equations of state exhibit approximately the
same angular distribution whereas the CASCADE calculation does not exhibit any
peak perpendicular to the event plane. For larger impact parameters, however,
also the CASCADE calculation shows a pronounced {\em squeeze--out}. }
\end{figure}

\begin{figure}
\caption{ \label{pisqb}{\em squeeze--out ratio} $R_{out/in}$ versus
impact parameter for neutral pions. The
calculations were performed with the
hard equation of state with (circles) and without
(squares) momentum dependence  as well as in CASCADE mode (triangles).
Cuts around mid-rapidity ($-0.2 \le y_{c.m.} \le 0.2$) and for high transverse
momentum ($p_t \ge 300$ MeV) were employed.
For large impact parameters the CASCADE
calculation agrees with the hard equation
of state without momentum dependence.
However, the onset of
{\em squeeze--out} in CASCADE mode is shifted towards larger
impact parameters in
comparison with the hard equation of state.
Most importantly, the inclusion of the momentum dependence results in
a drastic increase of the {\em squeeze--out ratio}.
The lines are inserted to guide the eye. }
\end{figure}


\begin{references}

\bibitem{sch68}
W.~Scheid, R.~Ligensa, and W.~Greiner.
\newblock Phys.~Rev.~Lett. {\bf 21}, 1479 (1968).

\bibitem{cse86}
L.~P.~Csernai and J.~I.~Kapusta.
\newblock Phys.~Reports~{\bf 131}, 225 (1986).

\bibitem{sto86}
R.~Stock.
\newblock Phys.~Reports~{\bf 135}, 261 (1986).

\bibitem{st86}
H.~St\"ocker and W.~Greiner.
\newblock Phys.~Reports~{\bf 137}, 277 (1986).

\bibitem{cl86}
R.~B.~Clare and D.~Strottman.
\newblock Phys.~Reports~{\bf 141}, 179 (1986).

\bibitem{schue87}
B.~Sch\"urmann, W.~Zwermann and R.~Malfliet.
\newblock Phys.~Reports~{\bf 147}, 3 (1986).

\bibitem{cas90}
W.~Cassing, V.~Metag, U.~Mosel and K.~Niita.
\newblock Phys.~Reports~{\bf 188}, 365 (1986).

\bibitem{st80}
H.~St\"ocker, J.~A.~Maruhn and W.~Greiner,
\newblock Phys.~Rev~Lett.~{\bf 44}, 725 (1980).

\bibitem{st82}
H.~St\"ocker, L.~P.~Csernai, G.~Graebner, G.~Buchwald, H.~Kruse, R.~Y.~Cusson,
J.~A. Maruhn and W.~Greiner,
\newblock Phys.~Rev~{\bf C25}, 1873 (1982).


\bibitem{st78}
H.~St\"ocker, W.~Greiner, and W.~Scheid.
Z.~Phys.~{\bf A286}, 121 (1978).

\bibitem{da79}
P.~Danielewicz.
Nucl.~Phys.~{\bf A314}, 465 (1979).

\bibitem{st81}
H. St\"ocker, A.~A.~Ogloblin and W.~Greiner.
\newblock Z.~Phys.~{\bf A303}, 259 (1981).

\bibitem{san80}
A.~Sandoval, R.~Stock, H.~E.~Stelzer, R.~E.~Renfordt, J.~W.~Harris,
J.~P.~Brannigian, J.~V.~Geaga, L.~J.~Rosenberg, L.~S.~Schroeder and
K.~L.~Wolf.
Phys.~Rev.~Lett. {\bf 45},874 (1980).

\bibitem{sto82}
R.~Stock, R.~Bock, R.~Brockmann, J.W.~Harris, A.~Sandoval, H.~Str\"obele,
                K.L.~Wolf, H.G.~Pugh, L.S.~Schroeder, M.~Maier, R.E.~Renfordt,
                A.~Dacal and M.E. Ortiz.
Phys.~Rev.~Lett.~{\bf 49}, 1236 (1982).

\bibitem{har85}
J.~Harris, R.~Bock, R.~Brockmann, A.~Sandoval, R.~Stock, H.~Stroebele,
  G.~Odyniec, L.~Schroeder, R.~E. Renfordt, D.~Schall, D.~Bangert, W.~Rauch,
  and K.~L. Wolf.
\newblock Phys.~Lett.~{\bf B153}, 377 (1985).

\bibitem{nag81}
S.~Nagamiya, M.C. Lemaire, E.~Moeller, S.~Schnetzer, G.~Shapiro, H.~Steiner,
and I.~Tanihata.
Phys.~Rev.~{\bf C24}, 971 (1981).

\bibitem{bert84a}
G.F.~Bertsch, H.~Kruse and S.~Das~Gupta.
\newblock Phys.~Rev.~{\bf C29}, R673 (1984).

\bibitem{kru85a}
H.~Kruse, B.~V. Jacak, and H.~St\"ocker.
\newblock  Phys.~Rev.~Lett.~{\bf 54}, 289 (1985).


\bibitem{shor}
A.~Shor et~al.
\newblock Phys.~Rev.~Lett.~{\bf 48}, 1597 (1982).

\bibitem{ai85b}
J.~Aichelin and C.~M.~Ko.
\newblock Phys.~Rev.~Lett.~{\bf 55}, 2661 (1985).

\bibitem{ai87b}
J.~Aichelin, A.~Rosenhauer, G.~Peilert, H.~St\"ocker, and W.~Greiner.
\newblock  Phys.~Rev.~Lett.~{\bf 58}, 1926 (1987).

\bibitem{berg94}
F.~D.~Berg and the TAPS collaboration.
\newblock Phys.~Rev.~Lett.~{\bf 72}, 977 (1994).

\bibitem{mis94}
D.~Miskowiec and the KaoS collaboration.
\newblock Phys.~Rev.~Lett.~{\bf 72}, 3650 (1994).

\bibitem{ha88}
Ch. Hartnack, H.~St\"ocker, and W.~Greiner.
\newblock In H.~Feldmeier, editor, {\em Proc. of the International Workshop on
  Gross Properties of Nuclei and Nuclear Excitation, XVI, Hirschegg,
  Kleinwalsertal, Austria} (1988).

\bibitem{lib91a}
B.~A. Li, W.~Bauer, and G.~F. Bertsch.
\newblock {\bf Phys.~Rev. C} 44, 2095 (1991).

\bibitem{baplb}
S.~A.~Bass, C.~Hartnack, R.~Mattiello, H.~St\"ocker and W.~Greiner.
Phys.~Lett.~{\bf B302}, 381 (1993).

\bibitem{tra94}
M.~Trzaska and the FOPI collaboration.
Proc. of the XXXII Winter Meeting on Nucl.~Physics, Bormio, Italy, Jan. 1994.

\bibitem{baprl}
S.~A.~Bass, C.~Hartnack, H.~St\"ocker and W.~Greiner.
Phys.~Rev.~Lett.~{\bf 71}, 1144 (1993).


\bibitem{brill93a}
D. Brill and the KaoS collaboration.
Phys.~Rev.~Lett.~{\bf 71}, 336 (1993).

\bibitem{ven93}
L. Venema and the TAPS collaboration.
Pys.~Rev.~Lett.~{\bf 71}, 835 (1993).

\bibitem{moli85b}
J.~J~Molitoris and H.~St\"ocker,
\newblock Phys.~Rev~{\bf C32}, R346 (1985).

\bibitem{ai85a}
J.~Aichelin and G.~Bertsch.
\newblock  Phys.~Rev.~{\bf C31}, 1730 (1985).


\bibitem{ue33}
E.~A. Uehling and G.~E. Uhlenbeck.
\newblock  Phys. Rev. {\bf 43}, 552 (1933) and
Phys. Rev. {\bf 44}, 917 (1934).


\bibitem{greg87}
C.~Gregoire, B.~Remaud, F.~Sebille, L.~Vinet, and Y.~Raffray,
Nucl. Phys. {\bf A465}, 317 (1987).

\bibitem{ai86}
J.~Aichelin and H.~St\"ocker.
\newblock  Phys.~Lett.~{\bf B176}, 14 (1986).

\bibitem{pei89}
G.~Peilert, H.~St\"ocker, A.~Rosenhauer, A.~Bohnet, J.~Aichelin and
W.~Greiner.
\newblock Phys.~Rev.~{\bf C39}, 1402 (1989).

\bibitem{ai91}
J.~Aichelin.
\newblock  Phys.~Reports~{\bf 202}, 233 (1991).

\bibitem{fmd}
H.~Feldmeier.
\newblock Nucl.~Phys.~{\bf A515}, 147 (1990).

\bibitem{amd}
A.~Ono, H.~Horiuchi, T.~Maruyama and A.~Ohnishi.
\newblock Phys.~Rev.~Lett.~{\bf 68}, 2898 (1992).

\bibitem{pqmd}
G.~Peilert, J.~Konopka, M.~Blann, M.~G.~Mustafa, H.~St\"ocker and W. Greiner.
\newblock Phys.~Rev.~{\bf C46}, 1457 (1992).

\bibitem{ha89}
C.~Hartnack, L.~Zhuxia, L.~Neise, G.~Peilert, A.~Rosenhauer, H.~Sorge,
  J.~Aichelin, H.~St\"ocker, and W.~Greiner.
\newblock Nucl.~Phys.~{\bf A495}, 303 (1989).

\bibitem{hart}
Ch. Hartnack.
\newblock PhD thesis, GSI-Report 93-5 (1993).

\bibitem{ar82}
L.~G.~Arnold et al.
\newblock Phys.~Rev.~{\bf C25}, 936 (1982).

\bibitem{pa67}
G. Passatore.
\newblock Nucl.~Phys. {\bf A95}, 694 (1967).

\bibitem{bert88b}
G.~F.~Bertsch and S.~Das~Gupta,
\newblock Phys.~Rep.~{\bf 160}, 189 (1988).


\bibitem{wo90}
G.~Wolf, G.~Batko, W.~Cassing, U.~Mosel, K.~Niita, and M.~Sch\"afer.
\newblock Nucl.~Phys.~{\bf A517}, 615 (1990).

\bibitem{lib91b}
B.~A. Li and W.~Bauer.
\newblock Phys.~Lett.~{\bf B252}, 335 (1991).\\
B.~A. Li, W.~Bauer, and G.~F. Bertsch.
\newblock  Phys.~Rev.~{\bf C44}, 450 (1991).

\bibitem{yar79}
Y.~Yariv and Z.~Frankel.
\newblock Phys.~Rev.~{\bf C20}, 2227 (1979).

\bibitem{cug80}
J.~Cugnon.
\newblock  Phys.~Rev.~{\bf C22}, 1885 (1980).

\bibitem{eng94}
A. Engel, W. Cassing, U. Mosel, M. Sch\"afer and Gy. Wolf.
\newblock Nucl.~Phys.~{\bf  A572}, 657 (1994).


\bibitem{vw82}
B.~J. VerWest and R.~A. Arndt.
\newblock Phys.~Rev.~{\bf C25}, 1979 (1982).

\bibitem{da91}
P.~Danielewicz and G.~F. Bertsch.
\newblock  Nucl.~Phys.~{\bf A533}, 712 (1991).

\bibitem{rand}
J. Randrup.
\newblock Nucl.~Phys.~{\bf A314}, 429 (1979).

\bibitem{cug81}
J.~Cugnon, T.~Mizutani and J.~Vandermeulen.
\newblock Nucl.~Phys.~{\bf A352}, 505 (1981).

\bibitem{hub94}
S.~Huber and J.~Aichelin.
\newblock Nucl.~Phys.~{\bf A573}, 587 (1994).

\bibitem{ehe93}
W.~Ehehalt, W.~Cassing, A.~Engel, U.~Mosel and Gy.~Wolf.
\newblock Phys.~Lett.~{\bf B298}, 31 (1993).

\bibitem{xio93}
L.~Xiong, C.~M.~Ko and V.~Koch.
\newblock Phys.~Rev.~{\bf C47}, 788 (1993).

\bibitem{cha73}
G.~F.~Chapline, M.~H.~Johnson, E.~Teller and M.~S.~Weiss.
Phys.~Rev.~{\bf D8}, 4302 (1973)

\bibitem{bog}
J. Boguta.
Phys.~Lett~{\bf B109}, 251 (1982). \\
J. Boguta and H. St\"ocker.
Phys.~Lett~{\bf B120}, 289 (1983).

\bibitem{wald}
B. Waldhauser, J.A. Maruhn, H. St\"ocker and W. Greiner.
Z.~Phys. {\bf A328}, 19 (1987).

\bibitem{metag}
V. Metag.
Prog.~Part.~Nucl.~Phys. {\bf 30}, 75 (1993).

\bibitem{ehehalt}
W.~Ehehalt, W.~Cassing, A.~Engel, U.~Mosel and G.~Wolf.
Pys.~Rev. {\bf C47}, R2467 (1993).

\bibitem{mhof}
M. Hofmann, R.~Mattiello, N.~S.~Amelin, M.~Berenguer, A.~Dumitru, A.~Jahns,
A.~v.~Keitz, Y.~P\"urs\"un, T.~Sch\"onfeld, C.~Spieles, L.~A.~Winckelmann,
H.~Sorge, J.~A.~Maruhn, H.~St\"ocker and W.~Greiner.
Nucl.~Phys.~{\bf A566}, 15c (1994).

\bibitem{fuji}
W.~Benenson et al..
Phys.~Rev.~Lett.~{\bf 43}, 683 (1979).

\bibitem{taps2}
O. Schwalb and the TAPS collaboration.
Phys.~Lett.~{\bf B321}, 20 (1994).


\bibitem{chm94}
Ch. M\"untz and the KaoS collaboration.
Z.~Phys.~{\bf A} in print.

\bibitem{ba94a}
S.~A.~Bass, C.~Hartnack, H.~St\"ocker and W.~Greiner.
\newblock Phys.~Rev.~{\bf C50}, 2167 (1994).

\bibitem{st79}
H.~St\"ocker, J.~A.~Maruhn and W.~Greiner.
\newblock Z.~Phys. {\bf A293}, 173 (1979).

\bibitem{cse82}
L.~P.~Csernai, W.~Greiner, H.~St\"ocker, I.~Tanihata, S.~Nagamiya and
J.~Knoll.
\newblock Phys.~Rev~{\bf C25}, 2482 (1982).

\bibitem{gut79}
H.~H.~Gutbrod, A.~M.~Poskanzer and H.~G.~Ritter.
\newblock Rep. Prog. Phys. {\bf 52}, 1267 (1989).

\bibitem{moli85}
J.~J. Molitoris and H.~St\"ocker,
\newblock Phys.~Lett.~{\bf B162}, 47 (1985).

\bibitem{gus84a}
H.-A. Gustafsson, H.~H. Gutbrod, B.~Kolb, H.~L\"ohner, B.~Ludewigt, A.~M.
  Poskanzer, T.~Renner, H.~Riedesel, H.~G. Ritter, A.~Warwick, F.~Weik, and
  H.~Wieman.
\newblock Phys.~Rev.~Lett.~{\bf 52}, 1590 (1984).

\bibitem{moli87}
J.~J. Molitoris, H.~St\"ocker, and B.~L. Winer.
\newblock  Phys.~Rev.~{\bf C36}, 220 (1986).

\bibitem{gos89}
J.~Gosset O.~Vallette, J.~P.~Alard, J.~Augerat, R.~Babinet, N.~Basid,
F.~Brochard, N.~De~Marco, P.~Dupieux, Z.~Fodor, L.~Faysse, P.~Gorodetzky,
M.~C.~Lemaire, D.~L'Hote, B.~Lucas, J.~Marroncle, G.~Montarou, M.~J.~Parizet,
J.~Poitou, C.~Racca, A.~Rahmani, W.~Schimmerling and Y.~Terrien.
\newblock Phys.~Rev.~Lett. {\bf 62}, 1251 (1989).\\
\newblock J.~Poitou and the DIOGENE collaboration.
\newblock {\em Proceedings of the International Workshop on Gross Properties
        of Nuclei and Nuclear Excitations} XVI, Hirschegg, Kleinwalsertal,
        Austria, 1989.


\bibitem{ba94b}
S. A. Bass, M. Hofmann, C. Hartnack, H. St\"ocker and W. Greiner.
Phys.~Lett.~{\bf B335} (1994) 289.

\bibitem{moli86}
J.~J.~Molitoris, H.~St\"ocker, H.~A.~Gustafsson, J.~Cugnon and D. L'H\^ote.
\newblock Phys.~Rev.~{\bf C33}, 867 (1986).


\bibitem{bu83}
G.~Buchwald, G.~Gr\"abner, J.~Theis, H.~St\"ocker, K.~Frankel, M.~Gyulassy,
J.~Maruhn and W.~Greiner.
\newblock Phys.~Rev.~{\bf C28}, 2349 (1983).

\bibitem{ha90}
Ch.~Hartnack, H.~St\"ocker and W.~Greiner.
\newblock Proc. of the Nato Adv.~Study~Inst.
on the Nucl. Equation of State (Pe\~{n}iscola, Spain),
Editors W.~Greiner and H.~St\"ocker, Plenum Press (1990).

\bibitem{ch92}
Ch.~Hartnack, M.~Berenguer, A.~Jahns, A~.v.~Keitz, R.~Mattiello,
A.~Rosenhauer,
J.~Schaffner, Th.~Sch\"onfeld, H.~Sorge,
L.~Winckelmann, H.~St\"ocker and W. Greiner.
Nucl.~Phys.~{\bf A538}, 53c (1992).

\bibitem{ch93}
Ch.~Hartnack, J.~Aichelin, H.~St\"ocker and W. Greiner.
\newblock Mod.~Phys.~Lett. in print.

\bibitem{ba94c}
S. A. Bass, C. Hartnack, H. St\"ocker and W. Greiner.
GSI-preprint 94-12, submitted Z.~Phys.~{\bf A}.

\bibitem{gut89}
H.~H. Gutbrod, K.~H. Kampert, B.~W. Kolb, A.~M. Poskanzer, H.~G. Ritter, and
  H.~R. Schmidt,
\newblock Phys.~Lett.~{\bf B216}, 267 (1989).

\bibitem{gut89b}
H.~H.~Gutbrod, A.~M.~Poskanzer and H.~G.~Ritter.
Rep.~Prog.~Phys. {\bf 52}, 1267 (1989).

\bibitem{leifels}
The LAND collaboration, Y.~Leifels et al.,
Phys.~Rev.~Lett.~{\bf 71}, 963 (1993).


\end{references}
\end{document}